\documentclass[aps,twocolumn,pra,superscriptaddress,showpacs,nofootinbib]{revtex4}
\usepackage{amsmath,amssymb}
\usepackage{color}
\usepackage[pdftex]{graphicx}
\usepackage{hyperref}
\usepackage{mathptmx}
\usepackage{bm}

\begin{document}


\newcommand{\be}{\begin{equation}}
\newcommand{\ee}{\end{equation}}
\newcommand{\R}[1]{\textcolor{red}{#1}}
\newcommand{\B}[1]{\textcolor{blue}{#1}}
\newcommand{\huan}[1]{\textcolor{cyan}{#1}}
\newcommand{\haixing}[1]{\textcolor{red}{#1}}
\newcommand{\figref}[1]{\figurename~\ref{#1}}


\title{Quantum noise of white light cavity using double-pumped gain medium}

\author{Yiqiu Ma}
\affiliation{School of Physics, University of Western Australia, WA 6009, Australia}
\author{Haixing Miao}
\affiliation{School of Physics and Astronomy, University of Birmingham, B15, 2TT, United Kingdom}
\author{Chunnong Zhao}
\affiliation{School of Physics, University of Western Australia, WA 6009, Australia}
\author{Yanbei Chen}
\affiliation{Theoretical Astrophysics 350-17, California Institute of Technology, Pasadena,
CA 91125, USA}


\begin{abstract}
Laser interferometric gravitational-wave interferometers implement Fabry-Perot cavities to increase their peak sensitivity. However, this is at cost of reducing their detection bandwidth, which origins from the propagation phase delay of the light. The ``white-light-cavity" idea, first proposed by Wicht {\it et al.} [Optics Communications {\bf 134}, 431 (1997)], is to circumvent this limitation by introducing anomalous dispersion, using double-pumped gain medium, to compensate for such phase delay. In this article, starting from the Hamiltonian of atom-light interaction, we apply the input-output formalism to evaluate the quantum noise of the system. We find that apart from the additional noise associated with the parametric amplification process noticed by others, the stability condition for the entire system poses an additional constraint. Through surveying the parameter regimes where  the gain medium remains stable (not lasing) and stationary, we find that there is no net enhancement of the shot-noise limited sensitivity. Therefore, other gain mediums or different parameter regimes shall be explored for realizing the white light cavity.
\end{abstract}

\maketitle


\section{Introduction}
Second-generation large-scale interferometric gravitational wave (GW) detectors, such as advanced LIGO\,\cite{Harry2010}, advanced VIRGO\,\cite{Virgo} and KAGRA\,\cite{Somiya2012}, are designed to operate at better sensitivity than the first generation detectors. This improvement in sensitivity comes from increase in the optical power and introduction of a signal recycling mirror (SRM) to the initial configuration\,\cite{Rana2014}. The SRM on the dark port forms a signal recycling cavity (SRC) with the input test mass mirror (ITM). The position of SRM determines the propagation phase of the signal light inside the SRC, and control of the SRM parameters allows for adjustments to the frequency response of the interferometer\,\cite{Chen2002a,Chen2002b}. Two typical operational modes are the \emph{signal recycling mode} and \emph{resonant sideband extraction mode} (RSE). The signal recycling mode enhances the sideband carrier of the GW signal inside the cavity, while the RSE mode increases the detection bandwidth which is the effective bandwidth of the combined SRC and arm cavity\,\cite{Mizuno1993}.

However, broadening the detection bandwidth in the RSE mode comes at the loss of the peak sensitivity; while enhancing the peak sensitivity in SRC results in a narrower detection bandwidth. This trad-off is represented by the integrated sensitivity:
\be\label{limit}
\rho=\int^{\omega_\text{FSR}}_0\frac{1}{S_{hh}(\Omega)}d\Omega
=\frac{2\pi L_{\text{arm}}P_c\omega_0}{\hbar c},
\ee
which only depends on the intra-cavity power $P_c$  and cavity length $L_{\text{arm}}$, and is independent of the property of SRC.
The $\omega_0, \omega_{\text{FSR}}, c$ here are the laser frequency, free-spectral range, and the speed of light, respectively. Here, we only consider the shot noise limited strain sensitivity $S_{hh}(\Omega)$ since radiation pressure noise can in principle be evaded using frequency dependent readout or sufficiently heavy test masses. Such a trade-off between bandwidth and peak sensitivity is due to the accumulated phase of the sideband field propagating inside the arm cavity. There are several proposals in the literature that try to achieve the resonant amplification of the signal without decreasing the bandwidth, using the idea of \emph{white light cavity}. Among those, Wicht {\it et al.} were the first to suggest placing an atomic gain medium with anomalous dispersion inside the SRC to cancel the propagation phase\,\cite{Wicht1997,Wicht2000}. This idea was then followed by Pati and Yum et al. with different types of active mediums\,\cite{Pati2007,Yum2013}.

\begin{figure}
  \includegraphics[width=0.48\textwidth]{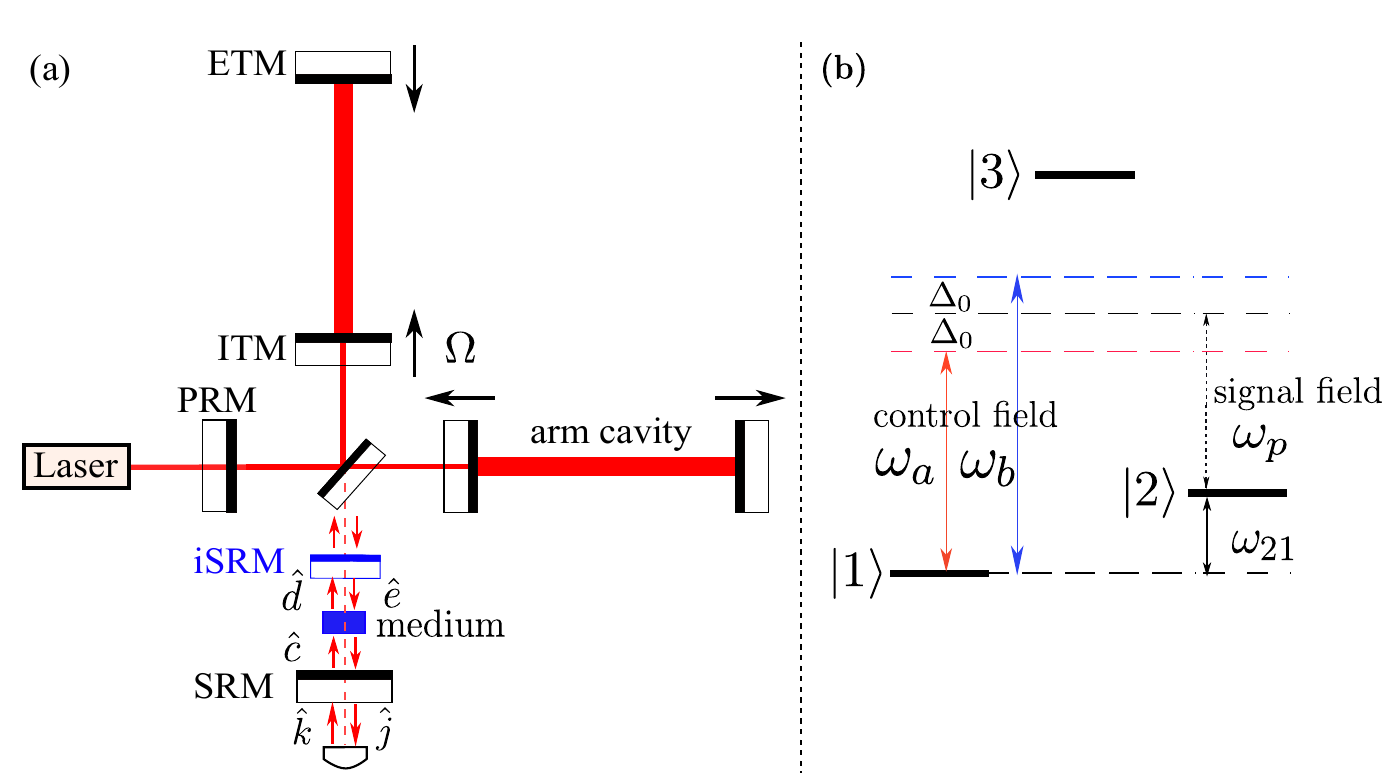}\\
  \caption{(a) the typical dual recycled interferometer configuration for an advanced gravitational wave detector, with an atomic gain medium (blue block) embedded inside the SRC to compensate the phase delay of the arm cavity. An internal SRM (iSRM) with the same transmissivity as ITM is introduced to make impedance matching so that effectively we can view the compound mirror (consists of ITM and iSRM) as transparent to the sideband field\,\cite{Salit}; (b) the energy levels of the gain medium atoms. Two far-detuned strong control laser with frequency $\omega_a$ and $\omega_b$ couple the energy levels $|3\rangle$ and $|1\rangle$.  The signal field interacts with $|2\rangle$ and $|3\rangle$.}\label{basic_scheme}
\end{figure}

The anomalous dispersion phenomenon and the interesting ``superluminal" physics of the propagation of light pulse in these active mediums have been theoretically discussed\,\cite{Dogariu2001} and experimentally demonstrated\,\cite{Akulshin1999,Wang2000}. In these experiments, the anomalous dispersion is usually realized by using a double-pumped gain medium in which the anomalous dispersion lies in between the two gain peaks. As discussed by  Kuzmich et.al\,\cite{Kuzmich2001}, the gain medium is subject to quantum noise that accompanies the amplification process. In addition, the gain medium could cause lasing when place inside a resonant cavity. To investigate how the quantum noise and associated gain influence the detector sensitivity and dynamics, we develop an input-output formalism for the optical field propagating through the gain medium. Using this formalism, we make a detailed analysis of the quantum shot noise limited sensitivity for a typical gravitational wave detector configuration implementing the white light cavity idea, as shown in ~\figref{basic_scheme}. Specifically, we consider: (i) the requirement for canceling the propagation phase shift; (ii) the optical stability of the interferometer system with the gain medium; (iii) the noise associated with the amplification process. Taking these factors into account, we find that the integrated shot noise limited sensitivity is still limited by Eq.\eqref{limit} when the gain medium itself is stable (not lasing) and stationary.

\section{A brief summary}
\begin{figure}
  \includegraphics[width=0.48\textwidth]{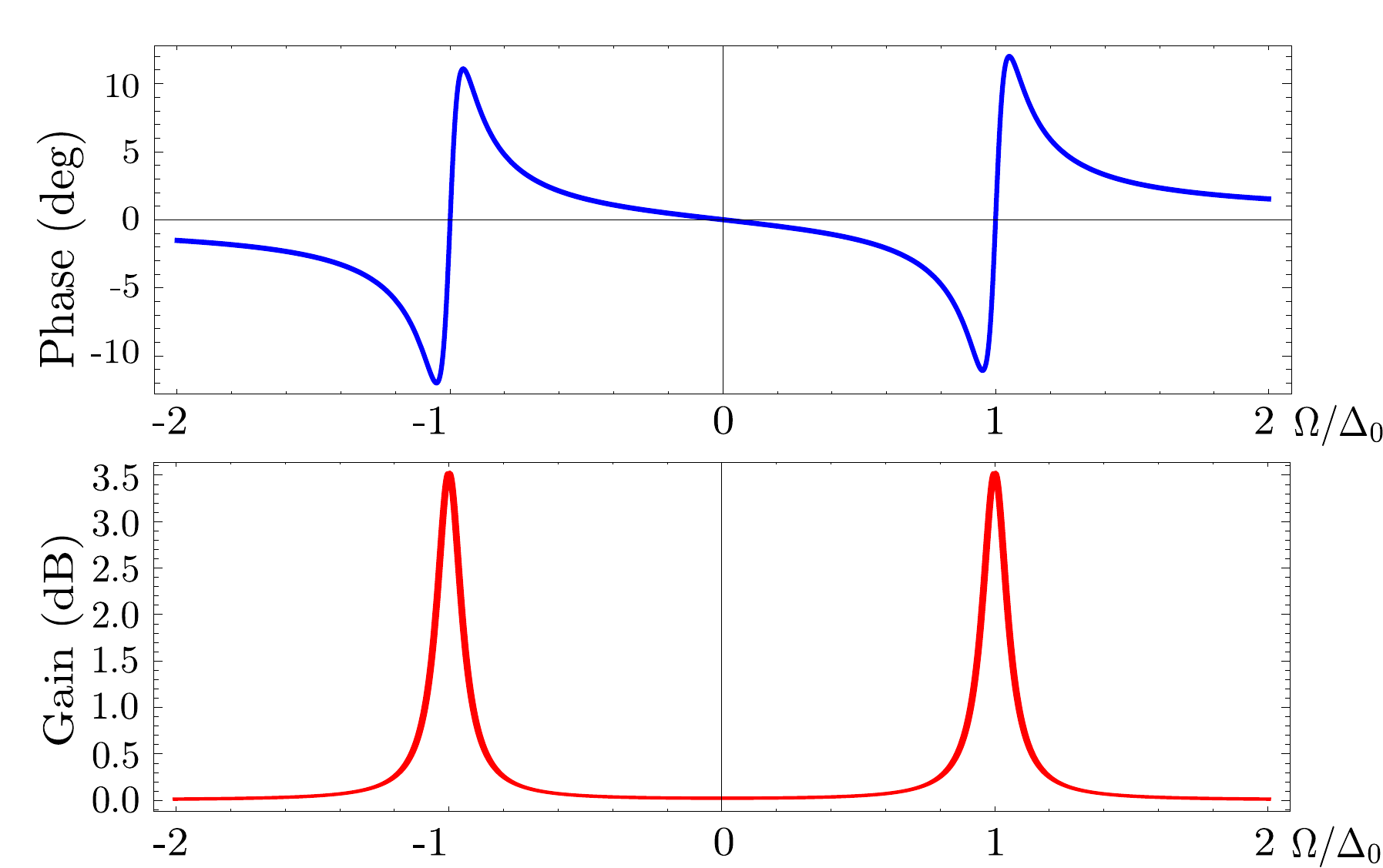}\\
  \caption{Phase angle and amplitude gain of the sideband field propagating through the atomic gain medium as functions of the normalized (by $\Delta_0$) sideband frequency. The top figure shows the negative dispersion of the atomic gain medium. The white light cavity bandwidth is the linear region between $-\Delta_0$ and $\Delta_0$. The bottom figure shows that the gain is negligibly small, except when $\Omega\sim \Delta_0$. In these frequency regions, the gain is high and need to be considered in the design for preventing the possible instability (See Section III for detailed analysis).}\label{dispersive}
\end{figure}

Before presenting the detailed analysis, we briefly summarize our main results in this section. The susceptibility of the double-pumped gain medium $\chi(\Omega)$ that we derive is given by (the same as in Refs.\,\cite{Wang2000,Dogariu2001} but with slightly different notations):
\be\label{suseptibility}
\chi(\Omega)=\frac{2i\Gamma_{\text{opt}}}{i(\Delta_0+\Omega)-\gamma_{12}+\Gamma_{\text{opt}}}
+\frac{2i\Gamma_{\text{opt}}}{i(-\Delta_0+\Omega)-\gamma_{12}+\Gamma_{\text{opt}}},
\ee
where $\Delta_0$ is one half of the frequency difference between two control fields, and $\Omega$ is the sideband frequency of the probe field with the carrier frequency $\omega_p$. The damping rate $\gamma_{12}$ is the effective atomic transition rate from state $|2\rangle$ to $|1\rangle$, while $\Gamma_{\text{opt}}$, which depends on pumping power of the control fields, is the transition rate between $|1\rangle$ to $|2\rangle$ mediated by a virtual excitation of $|3\rangle$. In terms of $\chi$, the ingoing
and outgoing field $\hat a_{\text{in}}$, $\hat a_{\text{out}}$ are related by (temporarily ignoring additional noise term that will be mentioned later):
\be\label{inoutsimplify}
\hat a_{\text{out}}(\Omega)=[1+i\chi(\Omega)/2]\hat a_{\text{in}}(\Omega).
\ee
In obtaining $\chi$, we use the following approximation:
\be\label{validycondition}
\Delta_0^2+(\gamma_{12}-\Gamma_{\text{opt}})^2\gg\Gamma_{\text{opt}}^2/4,
\ee
(For details, see Section III and Appendix B). Violation of this condition means the dynamics of the gain medium is non-stationary, as the two pumping fields start to interfere with each other.
This approximation is also equivalent to the weak-coupling approximation which allows us to express the input-output relation Eq.\eqref{inoutsimplify} for an unidirectional sideband field passing through the gain medium as:
\be\label{in-out1}
\hat a_{\text{out}}(\Omega)\approx e^{i\chi_r(\Omega)/2}e^{-\chi_i(\Omega)/2}\hat a_{\text{in}}(\Omega).
\ee
Here $\chi_r(\Omega)$ and $-\chi_i(\Omega)$ are the real and imaginary part of the susceptibility $\chi(\Omega)$ of the medium, which describe, respectively, the phase accumulation and the amplitude change of the sideband field after passing through the medium as shown in \figref{dispersive}.

In order to compensate the round-trip propagation phase inside the arm cavity thereby broadening the bandwidth of the optical cavity, the susceptibility should satisfy $d\chi(0)/d\Omega\approx -2L_{\text{arm}}/c$ (negative dispersion), which leads to:
\be\label{PCcondition}
\frac{\Gamma_{\text{opt}}[(\Gamma_{12}-\Gamma_{\text{opt}})^2-\Delta_0^2]}
{[(\Gamma_{12}-\Gamma_{\text{opt}})^2+\Delta_0^2]^2}=-\frac{L_{\text{arm}}}{c}.
\ee
Once we fixed the parameter $\Gamma_{\text{opt}}$ and $\gamma_{12}$, we will have a pair of roots for $\Delta_0^2$. For the positiveness of these $\Delta_0^2$, the following condition
has to be satisfied (see Appendix B for detailed derivation):
\be\label{phasecancellationcriteria}
(\gamma_{12}-\Gamma_{\text{opt}})^2<\Gamma_{\text{opt}}c/(8L_{\text{arm}}).
\ee

\begin{figure}
  \includegraphics[width=0.48\textwidth]{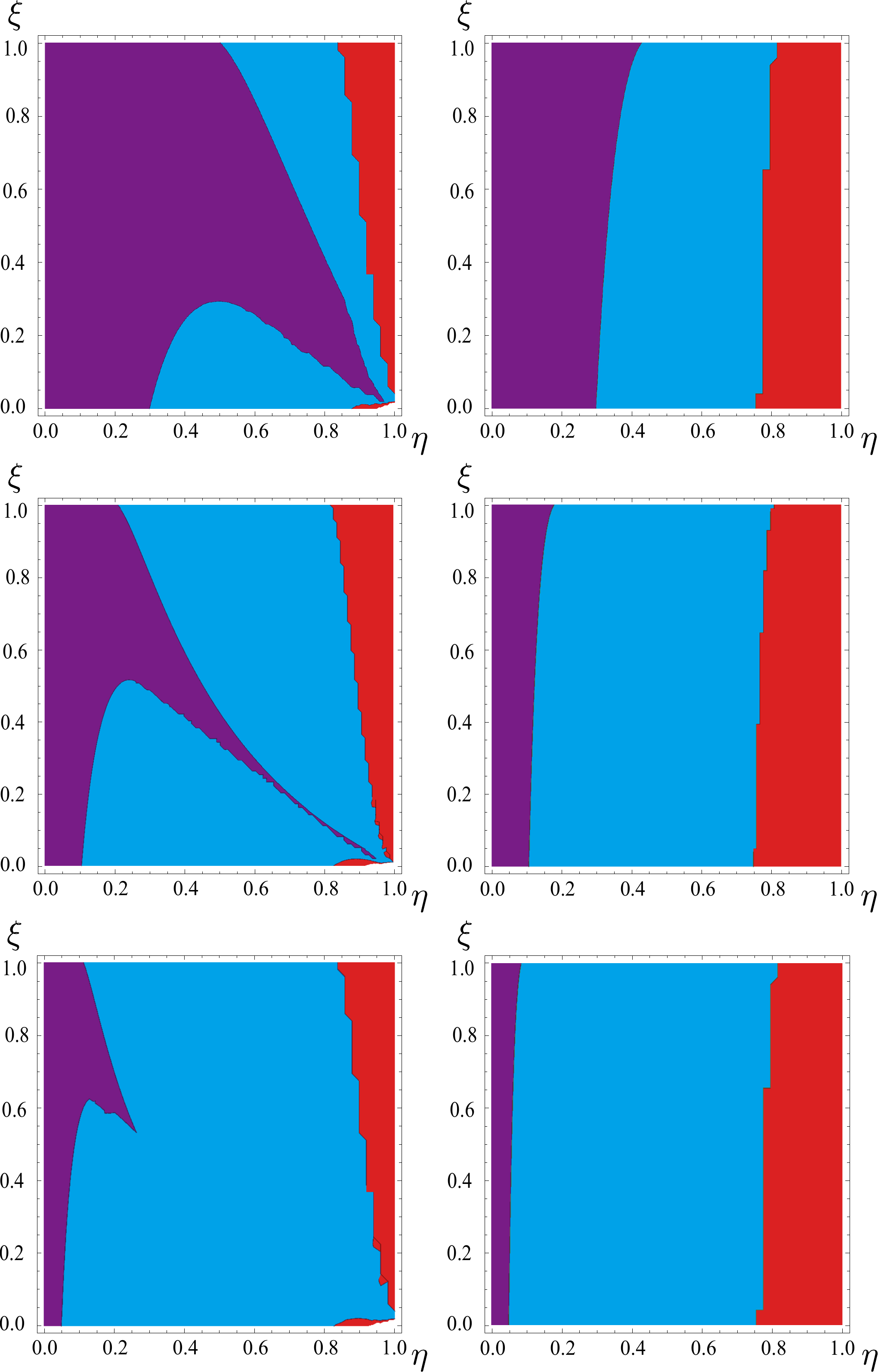}\\
  \caption{Stability region of the full interferometer scheme with double-pumped gain medium (optical stability only).  The SRM power reflectivity $r_s^2=0.5,0.8,0.9$ are chosen from the top panel to the bottom, while we survey the parameter region for $\Gamma_{\text{opt}},\gamma_{12}$. The horizontal and vertical axis are $\eta=\Gamma_{\text{opt}}/\gamma_{12}$ and $\xi=8(\gamma_{12}-\Gamma_{\text{opt}})^2L/c\gamma_{12}$. We survey $\eta,\xi$ between 0 and 1 so that atomic instability is excluded and the phase cancellation condition can be satisfied. For each $r_s$, the left panel and right panel correspond to two roots of $\Delta_0^2$ in Eq.\eqref{PCcondition}, respectively. The purple region is the only stable region. In the blue region (``optical instability region"), the atomic medium is stable by itself but the dynamics of the full interferometer system is unstable (see Section IV for details). The red region corresponds to the situation when the system becomes non-stationary, i.e. breaking down of condition in Eq.\eqref{validycondition}. With the increasing of the SRM reflectivity, the stable region shrinks due to the enhancement of the optical instability effect}\label{stability}
\end{figure}

Under these two conditions in Eq.\eqref{PCcondition} and Eq.\eqref{phasecancellationcriteria}, we explore the relevant parameter regime for studying the
dynamical behavior of the gain medium. Firstly, as we analyze in detail in Section.III and IV(A), the system has two different types of instability (lasing) 1) if $\gamma_{12}<\Gamma_{\text{opt}}$, there will be a population inversion between level $|1\rangle$ and $|2\rangle$, the gain medium starts lasing by itself, which we name as ``\textbf{atomic instability}";  2) if the photon loss rate for each round trip inside the cavity is less than the photon increasing rate through the amplification by the gain medium, the cavity-medium system starts lasing, which we name as ``\textbf{optical instability}". In Fig.\ref{stability}, we plot the phase diagram for the stability of the system. This figure gives a constraint on the possible parameter region for $\gamma_{12}$ and $\Gamma_{\text{opt}}$ of the atomic gain medium (with fixed SRM reflectivity $r_{s}$), if lasing were to be avoided.
Notice that we choose the re-scaled parameter $(\eta=\Gamma_{\text{opt}}/\gamma_{12},\xi=8(\gamma_{12}-\Gamma_{\text{opt}})^2L/c\gamma_{12})$ instead of $(\gamma_{12},\Gamma_{\text{opt}})$ and survey them within $0<\eta,\xi<1$. These new parameters help us exclude the atomic instability region ($\eta>1$) and the region where the phase-cancelation condition is unsatisfied ($\xi>1$).

\begin{figure}
  \includegraphics[width=0.48\textwidth]{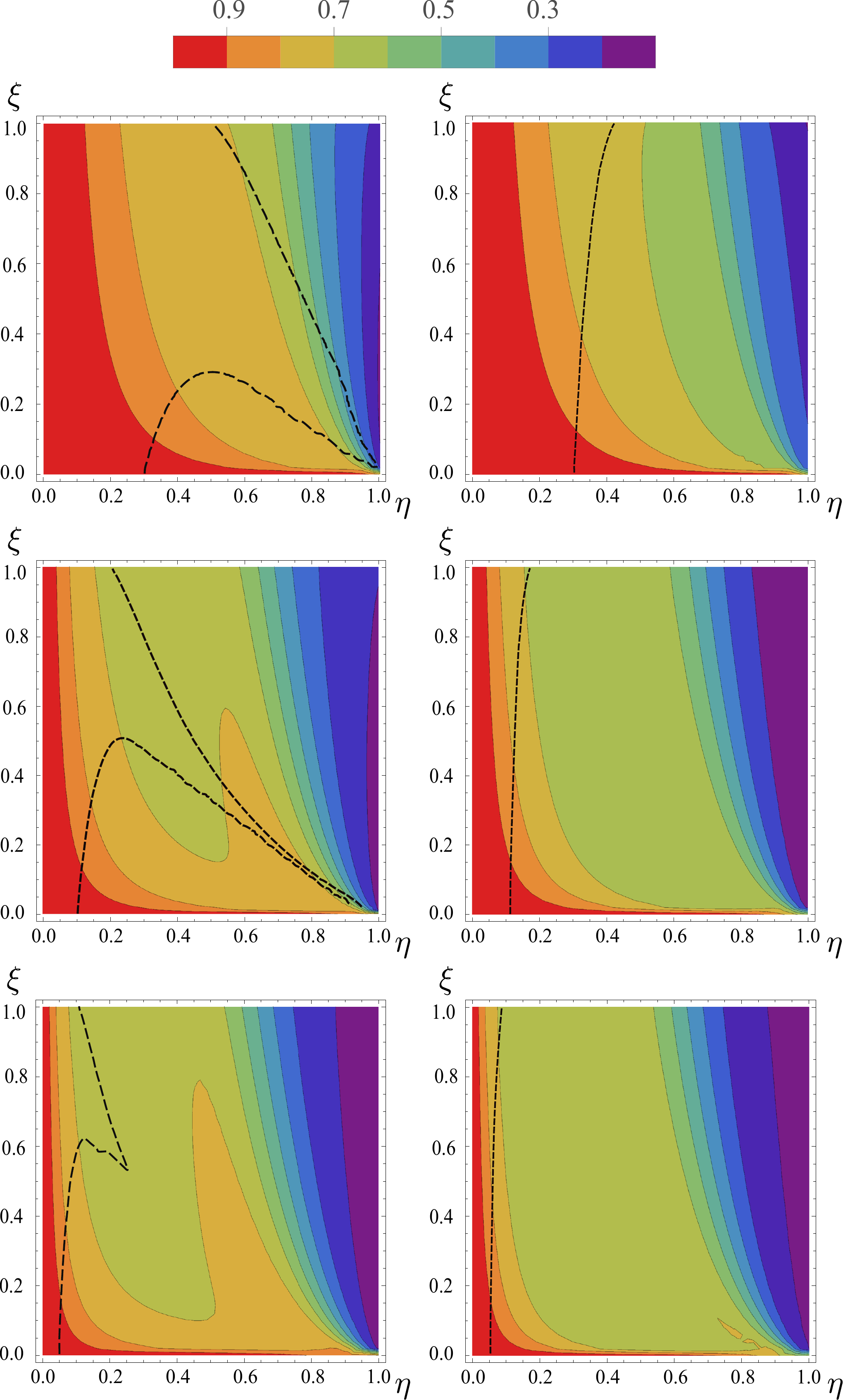}\\
  \caption{Integrated shot noise limited sensitivity improvement factor (defined in Eq.\eqref{SNRg}) of the full interferometer scheme with double-pumped gain medium, taking into account of the effect of additional noise. The specification for the parameters is identical to the one for producing \figref{stability}. The left panel and right panel correspond to the two roots of Eq.\eqref{PCcondition}. The dashed line is the boundary of the stable region shown in \figref{stability}. It is clear from this figure that there is no parameter region where the integrated shot noise limited sensitivity improvement factor is larger than 1, when the double-pumped gain medium itself is stable and stationary.}\label{gain_noise}
\end{figure}

Secondly, as implied by the above input-output relation, the gain medium is a parametric amplifier. Therefore, as first discussed by Caves\,\cite{Caves1982}, there must be an additional noise term on the right hand side of Eq.\eqref{in-out1} for keeping the commutation relation for $\hat a_{\text{out}}$ to be $[\hat a_{\text{out}}(t),\hat a^\dagger_{\text{out}}(t')]=\delta(t-t')$. This additional noise is due to the quantum fluctuation that causes spontaneous transition between $|1\rangle$ and $|2\rangle$. and degrades the signal to noise ratio. From the Hamiltonian, we can derive the noise terms from Heisenberg equations of motion. Their effect on the integrated shot noise limited sensitivity improvement factor (defined in Eq.\eqref{SNRg}) is given in \figref{gain_noise} (with tunable parameters of atomic system and fixed SRM reflectivity).

From these two figures, it is clear that 1) the stability condition and the phase cancelation condition put a strong constraint on the possible parameter region; 2) There is no parameter region where the shot-noise limited sensitivity is improved. This indicates that placing a stable double-pumped gain medium with anomalous dispersion inside the SRC can not broaden the detection bandwidth while increasing the shot-noise limited sensitivity. Therefore, one shall explore other types of gain mediums or different parameter regimes for realizing the white light cavity.

\section{Input-output relation of double gain atomic medium}
After summarizing the main results, we now start a detailed discussion by first developing an input-output formalism for light propagating through
the double-pumped gain medium in the Heisenberg picture.
As we have briefly mentioned in the \emph{Introduction}, our gain medium consists of three-level atoms schematically shown in ~\figref{basic_scheme}, with two red (blue)-detuned (with respect to frequency difference between $|3\rangle$ and $|1\rangle$) control lasers. The polarizations of the control and probe fields are orthogonal to each other and only sensitive to the atomic transitions between $|1\rangle\leftrightarrow|3\rangle$ and
$|2\rangle\leftrightarrow|3\rangle$, respectively. In modeling the gain medium, we treat the atoms as non-interacting distinguishable particles. Nevertheless, all the atoms have the same energy level structures. In this section, we first derive the atomic dynamics for a single three-level atom, then extend the result to many-atoms case under the approximation that the length of the gain medium is much smaller than the spatial scale of the optical sideband field $2\pi c/\Omega$ where $\Omega$ is the gravitational wave frequency.

\subsection{Single-atom dynamics}
The above physical modeling leads to the following system Hamiltonian for a single atom
\be\label{general Hamiltonian}
\hat H=\hat H_{\text{atom}}+\hat H_{\text{f}}+\hat H_{\text{int}}+\hat H_\gamma.
\ee
The $\hat H_{\text{atom}}$ is the free Hamiltonian for a three level atoms in the form of:
\be\label{single atom Hamiltonian}
\hat H_{\text{atom}}=\sum_{a=1,2,3}\hbar\omega_a\hat \sigma_{aa},
\ee
where $\omega_a$ is the Bohr frequency of energy level $a$ and
$\hat \sigma_{aa}$ is the  atomic population operator.

The $\hat H_{f}$ is the free Hamiltonian for the sideband probe fields propagating in the main GW detector. Since we are only interested about optical modes that are around the central frequency of the probe field $\omega_p$, these modes have frequency $\omega_p\pm\Omega$ where $\Omega$ denotes the frequency band that we are focusing on. Then we have the Hamiltonian $\hat H_{f}$ as:
\be
\begin{split}
\hat H_{f}&=\hbar c\int^{\Delta k}_{-\Delta k} dk'(k_p-k')\hat a_{-k_p+k'}^\dag\hat a_{-k_p+k'}\\
&\approx\hbar c\int^\infty_{-\infty}dk'(k_p-k')\hat a_{-k_p+k'}^\dag\hat a_{-k_p+k'},
\end{split}
\ee
where we have assumed that the field propagate uni-directionally (from right to left) and $k_p=\omega_p/c,\Delta k=\Omega/c$. We have also used the narrow band approximation $\Delta k/k_p=\Omega/\omega_p\ll1$ so that we can extend the integral range to $[-\infty,\infty]$.  By defining the optical creation/annihilation operators in $x-$domain as:
$\hat a_x=\int^\infty_{-\infty}dk'\hat a_{-k_p+k'}e^{-ik'x}$, the above Hamiltonian can be written as (for details, see\,\cite{Chen2013,Hong2013} or \emph{Appendix A}):
\be\label{free field Hamiltonian}
\hat H_{\text{f}}=\frac{i\hbar c}{2}\int^\infty_{-\infty}\left[\frac{\partial \hat a_x^\dagger}{\partial x}\hat a_x-\hat a_x^\dagger\frac{\partial \hat a_x}{\partial x}\right]dx.
\ee
Notice that the $\hat a_x$ is the slowly varying amplitude operator (both spatially and temporary) with respect to $e^{-i\omega_px/c-i\omega_pt}$, defined as: $\hat{E}_p(t,x)=\hat a_x(t)e^{-i\omega_px/c-i\omega_pt}+h.c$. The probe field propagates unidirectionally (right to left), encounters and interacts with the atom at position $x_0$. This interaction is given by a Jaynes-Cumming type of Hamiltonian under the rotating wave approximation\,\cite{Scully1997}:
\be\label{interaction H}
\begin{split}
\hat H_{\text{int}}=-\frac{\hbar}{2}\mu_{23}\hat a^\dagger_{x_0}e^{i\omega_pt}\hat \sigma_{23}-\frac{\hbar}{2}\mu_{13}E^*_c
\hat \sigma_{13}+h.c,
\end{split}
\ee
where the first term describes the atomic transition between $|2\rangle$ and $|3\rangle$ under the driving of probe fields with transition operator $\hat\sigma_{23}$ and the second term describes the atomic transition between $|1\rangle$ and $|3\rangle$ under the pumping of control fields with the transition operator $\hat \sigma_{13}$. The $E_c=E_ae^{i\omega_at}+E_be^{i\omega_bt}$ describes the two classical amplitude of control fields with frequency $\omega_{a,b}$ and the $\mu_{mn} (m,n=1,2,3)$ are the dipole moments of the atom. The atom transition operators satisfy
the algebra:
\be
\hat\sigma_{mn}\hat\sigma_{kl}=\hat\sigma_{ml}
\delta_{nk}\text{ ; }(\hat\sigma_{mn})^\dagger=\hat\sigma_{nm}.
\ee

The coupling between an atom with other bath sources at position $x_0$ is introduced phenomenologically by:
\be\label{bath}
\begin{split}
\hat H_{\gamma}=&i\hbar\sqrt{2\gamma_{12}}\hat n_{12}^\dagger e^{i\omega_{21}t}\hat\sigma_{12}
-\frac{\hbar}{2}\mu_{13}\hat a_{c}^\dag e^{i\omega_0t}\hat\sigma_{13}+h.c
\end{split}
\ee
where $\hat n_{12}$ is the noise operator which couples to the atomic transition operator between $|1\rangle$ and $|2\rangle$. The $\hat a_{c}$ is the quantum fluctuation associated with the control field. In our 1-D model, the quantum fluctuation associated with the probe field has been included in the $\hat a_{x_0}$ field of Eq.\eqref{interaction H}.  The noise bathes model can be attributed to mutual collision of atoms or the stimulation of the external electromagnetic vacuum bath. Here to study the minimal additional noise, we consider an effectively zero-temperature external bath.

With the Hamiltonian, we can now analyze the dynamics of the gain medium. The Heisenberg equation of motion for the probe field derived from Eq.\eqref{free field Hamiltonian} can be written as:
\be\label{fieldeom}
\begin{split}
&\frac{\partial\hat a_x}{\partial t}-c\frac{\partial\hat a_x}{\partial x}=\frac{i}{2}\mu_{23}\hat\sigma_{23}e^{-i\omega_{p}t}\delta(x-x_0),\\
\end{split}
\ee
which reflects the fact that the probe field propagates from right to left (unidirectional).

The Heisenberg equations of motion for the atomic transition operators of a single atom in this gain medium are given by:
\begin{subequations}\label{nonlinearequations}
\begin{align}
\dot{\hat\sigma}_{13}+(i&\omega_{31}+\gamma_{13})\hat\sigma_{13}=
i\sqrt{2\gamma_{13}}(\hat\sigma_{11}-\hat\sigma_{33})\hat a_{c\rm in}e^{-i\omega_{0}t}\nonumber\\
&+\frac{i}{2}\mu_{13}(\hat\sigma_{11}-\hat\sigma_{33})E_c
+\frac{i}{2}\mu_{23}\hat \sigma_{12}\hat a_{x_0}e^{-i\omega_{p}t}\\
\dot{\hat\sigma}_{23}+(i&\omega_{32}+\gamma_{23})\hat\sigma_{23}
=\frac{i}{2}\mu_{13}\hat\sigma_{21}(E_c+\hat a_ce^{-i\omega_{0}t})\nonumber\\
&+i\sqrt{2\gamma_{23}}(\hat\sigma_{22}-\hat\sigma_{33})\hat a_{\rm in}e^{-i\omega_{p}t}\\
\dot{\hat\sigma}_{12}+(i&\omega_{21}+\gamma_{12})\hat\sigma_{12}
=-\sqrt{2\gamma_{12}}(\hat\sigma_{11}-\hat\sigma_{22})
\hat n_{12\rm in}e^{-i\omega_{21}t}\nonumber\\
&\frac{i}{2}\mu_{23}\hat\sigma_{13}\hat a^{\dagger}_{x_0}e^{i\omega_pt}-\frac{i}{2}\mu_{13}\hat\sigma_{32}(E_c+\hat a_ce^{-i\omega_0t})
\end{align}
\end{subequations}
Notice that the $\hat n_{12\rm in}$, $\hat a_{\rm in}$ and $\hat a_{c\rm in}$
are the incoming noise fields, whose relations with the $\hat n_{12},\hat a, \hat a_c$ of Eq.\eqref{bath} are given in the way of Eq.\eqref{a-bsigma23}\eqref{junction}.
The $\gamma_{13}=\mu_{13}^2/8,\gamma^s_{23}=\mu_{23}^2/8$ can be derived from Eq.\eqref{interaction H}. Besides, the condition that the majority of atoms are initially prepared at $|1\rangle$ is set as an assumption. In a real experiment, this population preparation can be achieved through various methods such as introducing an additional optical pumping field\,\cite{Wang2000}.

It is clear that the above equations of motion are generally nonlinear. However the system dynamics can be simplified by exploring the linear regime where the scheme is proposed to operate. The simplification can be done using perturbative method by noticing that 1) the control fields have large detuning with respect to $\omega_{31}$ and therefore the population of atoms on $|3\rangle$ is still small compared to that on $|1\rangle$; 2) the transition between $|1\rangle-|3\rangle$ is much stronger than the transition between $|1\rangle-|2\rangle$ and $|2\rangle-|3\rangle$ since it is induced by strong control beams;
3) The probe field is very weak compared to the control field since it is around the quantum level.

There are three dimensionless expansion parameters in this system of equations of motion: $\epsilon\sim\mu_{mn}E_c/\Delta_0,\alpha\sim\mu_{mn}\hat a/\Delta_0$ and $\alpha\ll\epsilon\ll1$ (notice that the denominator could also be other frequency scales such as $\omega_{31}-\omega_{a,b}$, we choose the smallest one here for briefness).
Writing the $\hat \sigma_{13}$, $\hat \sigma_{23}$, $\hat \sigma_{12}$ in the rotating frame of $\omega_0=(\omega_a+\omega_b)/2$, $\omega_p$ and $\omega_0-\omega_p$ respectively,
the leading order ($\sim\epsilon$) of $\hat \sigma_{13}$ dynamics can be derived as:
\be\label{sigma13}
\dot{\bar\sigma}_{13}-i(\omega_0-\omega_{31}+i\gamma_{13})\bar\sigma_{13}=
\frac{i}{2}\mu_{13}\bar{\sigma}_{11}(E_a e^{-i\Delta_0 t}+E_be^{i\Delta_0 t}),
\ee
in which we can approximate the collective population operator on $|1\rangle$ as $\bar{\sigma}_{11}\approx 1$ and $\bar\sigma_{mn}$
is the mean value of $\sigma^i_{mn}$ averaged over all the atoms. The solution of Eq.\eqref{sigma13} is given by:
\be\label{sigma13bar}
\bar{\sigma}_{13}\approx\frac{1}{2}\frac{\mu_{13}E_ae^{-i\Delta_0 t}}{\omega_{31}-\omega_{a}}+\frac{1}{2}
\frac{\mu_{13}E_be^{i\Delta_0 t}}{\omega_{31}-\omega_{b}}.
\ee
Here, we have neglected the $\gamma_{13}$ which is assumed to be much smaller than the detuning: $\gamma_{13}/(\omega_{a,b}-\omega_{31})\ll\epsilon$.
In the same rotation frame, the leading order of the $\hat \sigma_{23}$ ($\sim\epsilon^2\alpha$) and $\hat \sigma_{12}$ ($\sim\epsilon\alpha$) dynamics can be written as:
\begin{subequations}\label{atomeom}
\begin{align}
&\dot{\hat\sigma}_{23}-i(\omega_p-\omega_{32}+i\gamma_{23})\hat\sigma_{23}
=\frac{i}{2}\mu_{13}\hat \sigma_{21}\tilde{E}_c,\\
&\dot{\hat\sigma}_{12}+\gamma_{12}\hat\sigma_{12}
=\frac{i}{2}\mu_{23}\hat a^\dagger_{x_0}\bar \sigma_{13}-\frac{i}{2}\mu_{13}\hat{\sigma}_{32}
\tilde{E}_c\nonumber\\&\qquad\qquad\qquad\qquad
-\sqrt{2\gamma_{12}}\hat n_{12\rm in},
\end{align}
\end{subequations}
where $\tilde{E}_c=E_ae^{-i\Delta_0t}+E_be^{i\Delta_0t}$ is the pumping field strength in the rotating frame of $\omega_0$ and we have used the fact that $\omega_0=\omega_p+\omega_{21}$ (See \figref{basic_scheme}). We also make use of the fact that $\bar\sigma^j_{11}=1$. In deriving Eq.\eqref{atomeom}, we also assume that system parameters satisfy: $\gamma_{23}/(\omega_p-\omega_{32})\ll\epsilon^2\alpha$.

For the probe field, we can integrate Eq.\eqref{fieldeom} around $x_0$ and obtain:
\be\label{a-bsigma23}
-\hat a_{x_0+}+\hat a_{x_0-}=\frac{i}{2}\mu_{23}\hat\sigma_{23},
\ee
in which the $\hat a_{x_0+}$ and $\hat a_{x_0-}$ are the
incoming and outgoing sidebands fields (respectively) defined
in the vicinity of the interaction point $x_0$ (in the following, we will use $\hat a_{\rm in}$ and $\hat a_{\rm out}$ to represent them, respectively). The probe field at $x_0$ is connected with these vicinity fields through the junction condition:
\be\label{junction}
\hat a_{x_0}=\frac{1}{2}(\hat a_{\rm in}+\hat a_{\rm out}).
\ee

The dynamics of $\hat \sigma_{32}$ can be obtained by solving Eq.\eqref{atomeom}(a), we have:
\be\label{sigma23}
\hat\sigma_{32}=\frac{\mu_{13}}{2}\frac{\tilde{E}^*_c}{\omega_{32}-\omega_p}
\hat\sigma_{12},
\ee
In deriving the above equation, we assume that $\gamma_{23}\ll\omega_{32}-\omega_p$ and make use of the fact that $\sigma_{12}$ is a slowly varying amplitude thereby $\dot{\hat\sigma}_{32}\approx0$ in Eq.\eqref{atomeom}.
Substituting the solution Eq.\eqref{sigma23} and Eq.\eqref{junction} into Eq.\eqref{atomeom}(b) and Eq.\eqref{a-bsigma23}, we can \textbf{adiabatically eliminate} the $\hat \sigma_{23}$ so that the Eq.\eqref{atomeom}(b) and Eq.\eqref{a-bsigma23} form a closed equation set:
\begin{subequations}\label{a-bsigma12}
\begin{align}
&\hat a^\dag_{\rm out}=\hat a^\dag_{\rm in}-i(\sqrt{2\gamma_{\rm opta}}e^{i\Delta_0t}+
\sqrt{2\gamma_{\rm optb}}e^{-i\Delta_0t})\hat
\sigma_{12}\\
&\dot{\hat\sigma}_{12}+\gamma_{12}\hat\sigma_{12}=i\left(\sqrt{2\gamma_{\rm optb}}e^{-i\Delta_0t}+\sqrt{2\gamma_{\rm opta}}
e^{i\Delta_0t}
\right)\hat a_{\rm in}^\dag
\nonumber\\
&+(\gamma_{\text{opt}}+\gamma_{2\Delta_0})\hat\sigma_{12}-i\omega_{\text{opt}}
\hat\sigma_{12}-\sqrt{2\gamma_{12}}\hat n_{12\rm in}.
\end{align}
\end{subequations}
This equation set describes the coupling between the composite system (consisting of the atom and the pumping fields) to the probe field.

The second term on the right-hand-side of Eq.\eqref{a-bsigma12} is the sum of
an anti-damping term $\gamma_{\text{opt}}\hat \sigma_{12}$:
\be\label{gammaopt}
\gamma_{\text{opt}}=\gamma_{\text{opt}a}+\gamma_{\text{opt}b}
=\frac{\mu_{23}^2\mu_{13}^2}{32}\left[
\frac{|E_a|^2}{\Delta_a^2}+
\frac{|E_b|^2}{\Delta_a^2}\right],
\ee
and a high-oscillating term $\gamma_{2\Delta_0}\hat \sigma_{12}$:
\be
\gamma_{2\Delta_0}=\frac{\mu_{23}^2\mu_{13}^2}{32}\left[
\frac{E_aE^*_be^{2i\Delta_0 t}}{\Delta_a^2}+
\frac{E_bE^*_ae^{-2i\Delta_0t}}{\Delta_a^2}\right].
\ee
These formulae are derived under the approximation $\Delta_0\ll \Delta_{p,a,b}$ (thereby $\Delta_p\approx\Delta_a\approx\Delta_b$). These are good approximations to the situation in the proposed experiments \,\cite{Dogariu2001,Wang2000,Pati2007}.

In the symmetric pumping case when $E_a=E_b=E_0$:
\be
\gamma_{\text{opt}}\approx \mu_{23}^2\mu_{13}^2E_0^2/(16\Delta_a^2).
\ee
When $\gamma_{\text{opt}}$ is larger than $\gamma_{12}$, we have the population inversion between energy level $|1\rangle$ and $|2\rangle$, i.e., the atomic instability, mentioned earlier.

Solving the Eq.\eqref{a-bsigma12}(a)(b) in the frequency domain, we can obtain
the input-output relation for the probe field:
\be\label{in-out}
\begin{split}
\hat a_{\text{out}}(\Omega)=&\mathcal{M}(\Omega)\hat a_{\text{in}}(\Omega)+\mathcal{N}_{+}(\Omega)\hat n^{\dagger}_{12\rm in}(\Delta_0-\Omega)\\
&+\mathcal{N}_{-}(\Omega)\hat n^{\dagger}_{12\rm in}(-\Delta_0-\Omega),
\end{split}
\ee
with $\mathcal{M}(\Omega)$ and $\mathcal{N}(\Omega)$ given by:
\begin{subequations}
\begin{align}\label{trans}
&\mathcal{M}(\Omega)=1-\frac{\gamma_{\rm opt}}{i(\Omega+\Delta_0)-\gamma_{12}+\gamma_{\rm opt}}
-\frac{\gamma_{\rm opt}}{i(\Omega-\Delta_0)-\gamma_{12}+\gamma_{\rm opt}}\\\label{ncoefficients}
&\qquad\qquad \mathcal{N}_{\pm}(\Omega)=\frac{\sqrt{2\gamma_{12}\gamma_{\text{opt}}}}{\pm i\Delta_0-i\Omega+\gamma_{12}-\gamma_{\text{opt}}}.
\end{align}
\end{subequations}
Notice that 1) $\mathcal{N}^*_{\pm}(-\Omega)=\mathcal{N}_{\mp}(\Omega)$. 2) Here and after, for simplicity, we will only consider the symmetric pumping case where $E_a=E_b$ because the non-symmetric pumping will only induce an additional rotation int he quadrature plane, which does not affect our main results. 3) In obtaining the required susceptibility Eq.\eqref{suseptibility} and the corresponding input-output relation, we neglect the above oscillating terms at $2\Delta_0$ which is contributed by the driving from the beating of two control fields. This approximation leads to the condition in Eq.\eqref{PCcondition} (Details are discussed in Appendix B); 2) The tiny Stark frequency shift $\omega_{\text{opt}}=\mu_{13}^2|E_c(t)|^2/(4\Delta_p)\ll\Delta_0$
on the right hand side of Eq.\eqref{a-bsigma12} has been neglect here.

The above input-output relation describes a phase-insensitive parametric amplification process. Therefore, there is an additional noise given by the $\hat n^\dag_{12\rm in}$ terms in Eq.\eqref{in-out}. This noise comes from the stochastic dynamics of $\hat \sigma_{12}$ driven by the last term on the right hand side of Eq.\eqref{a-bsigma12}. The usual method for introducing the additional noise for parametric amplifier is using the argument given by Caves\,\cite{Caves1982}, which base on the principle that $\hat a_{\text{out}}$ field should satisfy Bosonic commutation relation. However, Caves's method can not be directly applied to our system since the additional noise has two different frequency channels. Solving the dynamics from the full Hamiltonian of the system can allows us to pin down the source of the additional noise and give the correct formula for the noise contribution.

\subsection{Extension to many-atoms case}
In the above subsection, we discussed the input-output relation for the probe field interacting with a single atom.
In this subsection, we extend the above results to the many-atoms case.

Since the size of the gain medium (centimeter scale) is much smaller than the spatial scale of the slowly varying amplitude of the probe field (kilometer scale),  therefore the slowly changing amplitude of the probe field interacts with all the atoms together. In this case, the anti-damping rate will be enhanced by a factor of $N$ where $N$ is the number of the atoms\,\cite{Polder1979,Dicke1954,Chiao1995} and the $\mathcal{M}$ coefficients in the above input-output relation can be written as:
\begin{subequations}\label{Mcoefficient}
\begin{align}\label{trans}
&\mathcal{M}(\Omega)=1-\frac{\Gamma_{\rm opt}}{i(\Omega+\Delta_0)-\gamma_{12}+\Gamma_{\rm opt}}
-\frac{\Gamma_{\rm opt}}{i(\Omega-\Delta_0)-\gamma_{12}+\Gamma_{\rm opt}}
\end{align}
\end{subequations}
where $\Gamma_{\rm opt}=N\gamma_{\rm opt}$.

The formulation of noise field in the input-output relation for many-atoms case has some subtleties, depends on the specific modeling of the interaction between the noise field and the atoms.

$\bullet$ \textbf{Noise interacts with atoms locally---}In this case, each atom is associated with its own noise bath. The noise term will be represented by:
\be\label{Nlocalcoefficient}
\text{Noise term}=\sum_{\pm}\sum^N_{j=1}\mathcal{N}_{\pm}(\Omega)
\hat n^{j\dag}_{12\rm in}(\pm\Delta_0-\Omega),
\ee
where
\be
\mathcal{N}_{\pm}(\Omega)=\frac{\sqrt{2\gamma_{12}\gamma_{\rm opt}}}{\pm i\Delta_0-i\Omega+\gamma_{12}-\Gamma_{\rm opt}}.
\ee

$\bullet$ \textbf{Noise interacts with atoms collectively---}
In some cases, the noise is introduced through processes where the electromagnetic field amplitude interacts with all the atoms collectively as what the slowly-varying amplitude of the probe field does. For example, the $\gamma_{12}$ is induced by applying an additional pumping laser such as the experiment done in\,\cite{Pati2007}. In this situation, the noise term will be represented by:
\be\label{Ncollectivecoefficient}
\text{Noise term}=\sum_{\pm}\mathcal{N}^c_{\pm}(\Omega)\hat n^\dag_{12\rm in}(\omega_{21}\pm\Delta_0-\Omega),
\ee
with
\be
\mathcal{N}^c_{\pm}(\Omega)=\frac{\sqrt{2\gamma_{12}\Gamma_{\rm opt}}}{\pm i\Delta_0-i\Omega+\gamma_{12}-\Gamma_{\rm opt}}.
\ee
Notice that the $\gamma_{12}$ here (also accordingly in $\mathcal{M}(\Omega)$) should be understood as $N$ times of the transition rate from $|2\rangle$ to $|1\rangle$ for one single atom which is proportional to the intensity of the additional pumping laser the experiment in\,\cite{Pati2007}.

Notice that 1): The input-output relations based on both of these noise models satisfies the Bosonic commutation relation  $[\hat a_{\rm out}(\Omega),\hat a_{\rm out}(\Omega')]=\delta(\Omega-\Omega')$ under the weak coupling approximation (See \emph{Appendix B}). For the single-pumping case where $\Delta_0=0$, the Bosonic commutation relation will be exactly satisfied.
2): More importantly, as we shall see later, the subtleties of the noise model \emph{do not }affect the sensitivity of the gravitational wave detector.

\subsection{Some physical discussion}
After deriving the system dynamics and input-output relation, we are going to give some intuitive discussion of the system dynamics and the additional noise.

Firstly, the ``anti-damping" dynamics of $\hat\sigma_{12}$ can be understood in the following way:  a small amount of atoms initially populated on $|1\rangle$ can be pumped to $|3\rangle$ by the detuned control fields, then jump to $|2\rangle$ due to their interactions with the probe field. During this indirect transition between $|1\rangle$ and $|2\rangle$ mediated by $|3\rangle$, the population of atoms on $|2\rangle$ will increase indefinitely if the decay rate from $|2\rangle$ to $|1\rangle$ is not sufficiently large--a ``\textbf{population inversion process}". Physically, this process could cause \textbf{lasing} (``atomic instability" in Section II ) and our approximation will fail as the population on $|2\rangle$ becomes larger than the population on $|1\rangle$. One may argue that this instability will not happen in real experiments with the atom population being prepared using additional pumping fields. However, the thermal collision relaxation rate can be tuned to be small if we decrease the gas temperature, increase the pumping beam waist and fill in the ``buffer" gas\,\cite{Bernheim1965,Dogariu2001}. In this case a small transition rate contributed by the optical pumping beam could be sufficient for the population preparation. Therefore, in principle the lasing can still happen as long the control fields are strong enough and $\Gamma_{\rm opt}>\gamma_{12}$.

Secondly, for the additional noise, the stochastic dynamics driven by the $\hat n_{12\rm in}$ can be attributable to 1) the collision of atoms due to the Van-der-Waals mutual interaction or the thermal collsion\,\cite{Dogariu2001}, 2) the transition between $|2\rangle-|1\rangle$ induced by environmental black-body radiation, 3) the contribution of the quantum noise associated with the additional optical pumping process as in\,\cite{Pati2007}.  In Eq.\eqref{in-out} and Eq.\eqref{Mcoefficient}-\eqref{Ncollectivecoefficient}, the $\hat a_{\text{out}}$ field contains terms related to the additional noise $\hat n$ in the way that the stochastic fluctuations of the population on $|1\rangle$ and $|2\rangle$ will cause fluctuations of the transition between $|2\rangle$ and $|3\rangle$, since $\hat\sigma_{23}$ is slaved by $\hat \sigma_{12}$.

In this section, we have derived the input-output relation for the sideband probe field propagating through the double gain medium from the full Hamiltonian. We also discussed the opto-atom dynamics and origin of the additional noise. In the next section, we will apply these results to the interferometer configuration shown in \figref{basic_scheme} and analyze its strain sensitivity.


\section{Interferometer with gain medium}
\begin{figure}
  \includegraphics[width=0.4\textwidth]{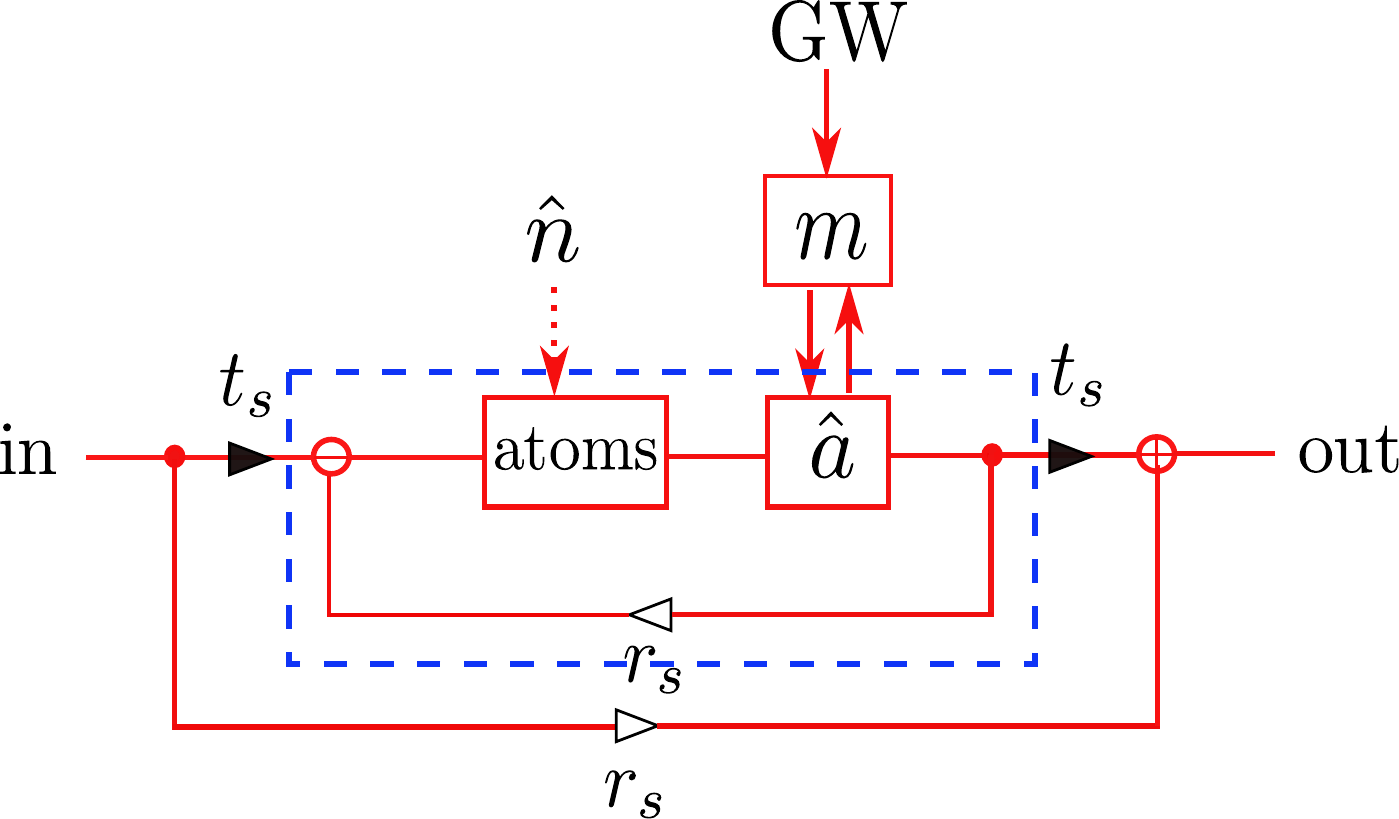}\\
  \caption{A flow chart showing the propagation of fields in the full system. The optical stability of the system is determined by the part inside the blue dashed box, whose gain function is given by Eq.\eqref{open loop gain} and \eqref{close loop gain}. The test mass (end mirror) $m$ is driven by gravitational waves while the double gain medium is affected by the additional noise $\hat n$. }\label{flow chart}
\end{figure}

The propagation of sideband fields inside the interferometer shown in \figref{basic_scheme} can be schematically
described by the flow chart shown in \figref{flow chart}.
Here, we only study the differential mode of this interferometer which carries the gravitational wave signal and can be mapped into a signal cavity containing a gain medium\,\cite{Chen2003}.
In this scheme, an internal signal recycling mirror is used to effectively remove the frequency response of the arm cavities so that the sideband fields see an input-output relation for the
amplitude and phase quadrature of sideband field in the following form (we ignore the optomechanical back-action term by assuming the infinitely heavy test masses):
\be
\bm e(\Omega)=\textbf{M}_0(\Omega)\bm d(\Omega)+\bm D (\Omega) h(\Omega),
\ee
or more explicitly, we can expand the vectors $\bm d,\bm e, \bm D$ and matrix $\textbf{M}_0$
\be
\left[
\begin{array}{c}
\hat e_1(\Omega)\\
\hat e_2(\Omega)
\end{array}
\right]
=e^{2i\Omega\tau}\left[
\begin{array}{cc}
1&0\\
0&1
\end{array}
\right]
\left[
\begin{array}{c}
\hat d_1(\Omega)\\
\hat d_2(\Omega)
\end{array}
\right]
+e^{i\Omega\tau}
\left[
\begin{array}{c}
0\\
\sqrt{2\mathcal{K}}
\end{array}
\right]h(\Omega)
\ee
with $\mathcal{K}=P_c\omega_0L^2/(\hbar c^2)$.

(A) In case of local noise model, the input-output relation for phase and amplitude quadrature of light field propagates through the gain medium is (Ref. Eq.\eqref{Mcoefficient}-\eqref{Ncollectivecoefficient}):
\begin{equation}\label{in-out quadrature}
\bm d(\Omega)
=M(\Omega)\bm c(\Omega)+\bm N_+(\Omega)+\bm N_{-}(\Omega).
\end{equation}
where $\bm N_{\pm}$ represent the additional noise terms, the forms of which depends on the specific noise modeling as we will show later.

The combined effect of the gain medium and the main interferometer is described by $\textbf{M}_{\rm tot}(\Omega)=M(\Omega)\textbf{M}_0(\Omega)$ with the noise terms $\bm{N}_{\pm}(\Omega)\textbf{M}_0(\Omega)$. Then the final input-output relation for the sideband field is given by:
\be\label{in-out detector}
\begin{split}
&\bm j(\Omega)=\textbf{M}_k(\Omega).
\bm k(\Omega)+t_{s}e^{i\Omega\tau}\textbf{M}_c(\Omega).\textbf{D}(\Omega)h(\Omega) \\&+t_{s}e^{2i\Omega\tau}\textbf{M}_c(\Omega).(\bm N_+(\Omega)+\bm N_-(\Omega)),\\
\end{split}
\ee
in which $\textbf{M}_k(\Omega)=-r_s\textbf{I}+t_{s}^2\textbf{M}_c(\Omega)\textbf{M}_{\rm tot}(\Omega)$ and $\textbf{M}_c(\Omega)=[\textbf{I}-r_{\rm s}\textbf{M}_{\rm tot}(\Omega)]^{-1}$ with $r_{s},t_s$ the amplitude reflectivity and transmissivity of the signal recycling mirror. The $\bm k(\Omega)$ and $\bm j(\Omega)$ are the input and output field of the entire configuration as shown in Fig.\ref{basic_scheme}.
The first term in Eq.\eqref{in-out detector} is the quantum noise contributed by the vacuum injection outside of the SRM, the second term is the contribution of additional noise introduced by the gain medium and the last term is the signal term. The SRM feeds the optical field back into the gain medium and the main interferometer, which is described by dashed blue box in the flow chart \figref{flow chart}. This feedback process will bring in another potential lasing (the ``optical instability" mentioned in Section II) even if $\gamma_{12}>\gamma_{\rm opt}$, discussed in the following subsection A.

For the additional noise terms $\bm N_{\pm}(\Omega)$ in the above relations, suppose the additional noise interacts with atoms locally, they are given by
\be
\bm N_+(\Omega)+\bm N_-(\Omega)=\sum^N_{j=1} [\textbf{N}_{\pm}(\Omega).\bm n^{j\pm}(\Omega)].
\ee
The $\bm n^{j\pm}$ are the vectors $(\hat n^{j\pm}_1,-\hat n^{j\pm}_2)$, with $\hat n^{j+(-)}_{1(2)}$ is the amplitude (phase)
quadrature of the noise field $\hat n^j_{12\rm in}$ with respect to the central frequency $\pm\Delta_0$. The noise matrix
$\textbf{N}_{\pm}$ can be derived using the sideband-quadrature transfer matrix $\textbf{M}_{qs}$ defined as:
\be
\textbf{M}_{\rm qs}=\frac{1}{\sqrt{2}}\left(
\begin{array}{cc}
1&1\\
-i&i
\end{array}
\right),
\ee
which leads to:
\be\label{nquadrature}
\textbf{N}_{\pm}(\Omega)=
\textbf{M}_{\rm qs}.
\left(
\begin{array}{cc}
\mathcal{N}_{\pm}(\Omega)&0\\
0&\mathcal{N}_{\mp}(\Omega)
\end{array}
\right).\textbf{M}_{\rm qs}^{-1}
\ee
in obtaining the above formula, we have made use of the relation: $\mathcal{N}^*_{+}(-\Omega)=\mathcal{N}_{-}(\Omega)$.

If the noise interacts with the atoms collectively, the noise term is given by:
\be
\bm N_+(\Omega)+\bm N_-(\Omega)=\sum_{\pm}
\textbf{M}_{qs}.
\left(
\begin{array}{cc}
\mathcal{N}^c_{\mp}(\Omega)&0\\
0&\mathcal{N}^c_{\pm}(\Omega)
\end{array}
\right).
\textbf{M}^{-1}_{qs}\bm n^{\pm}(\Omega)
\ee
where formally we have:
\be
\begin{split}
\mathcal{N}_{\pm}(\Omega)
=\sqrt{N}\frac{\sqrt{2\gamma_{12}\gamma_{\rm opt}}}{\pm i\Delta_0-i\Omega+\gamma_{12}-\Gamma_{\rm opt}}
\end{split}
\ee
where we have used the fact that $\Gamma_{\rm opt}=N\gamma_{\rm opt}$ while as before, the $\gamma_{12}$ here should be understood as N times of the $|2\rangle\rightarrow |1\rangle$ transition rate for a single atom. When we calculate the sensitivity, the $\sqrt{N}$ in the above equation will give us the same $N$ factor in the following Eq.\eqref{sensitivity} while the rest part of the above equation has the same form as $\mathcal{N}_{\pm j}$. Therefore if the the noise is modeled as interacting with the atoms collectively, the strain sensitivity formula will have the same form except that the $\gamma_{12}$ are different. However, since we will scan over all possible values of $\gamma_{12}$ and $\Gamma_{\rm opt}$ in the sensitivity calculation, this difference of $\gamma_{12}$ will not affect our final result.

Finally, the shot-noise limited strain sensitivity of the interferometer derived from Eq.\eqref{sensitivity}, which is given by:
\be\label{sensitivity}
\begin{split}
S^a_{hh}(\Omega)=&\frac{{\bm v}_h.\textbf{M}_k(\Omega).\textbf{M}_k^\dag(\Omega){\bm v}_h^T}{|t_s{\bm v}_h.\textbf{M}_c(\Omega).\textbf{D}(\Omega)|^2}\\
&+N\sum_{\pm}\frac{\bm v_h\textbf{M}_c(\Omega).\textbf{N}_{j\pm}(\Omega).
\textbf{N}_{j\pm}^\dag(\Omega)\textbf{M}_c^\dag(\Omega)\bm v_h^T}{|{\bm v}_h.\textbf{M}_c(\Omega).\textbf{D}(\Omega)|^2},
\end{split}
\ee
in which the $\bm v_h=(\sin\xi,\cos\xi)$ is the homodyne readout vector with $\xi$ is the homodyne angle. In our calculation, we will choose $\xi=0$ which means we only measure the phase quadrature $j_2(\Omega)$ of the output field.

We will discuss the numerical result of this shot-noise limited strain sensitivity in the following subsection B.

\subsection{Stability Criteria}

Besides the atomic instability due to ``population-inversion process" when $\gamma_{12}<\Gamma_{\text{opt}}$, it is important to notice that the dynamics of the interferometer with the gain medium may still be unstable (start lasing) even if $\gamma_{12}>\Gamma_{\text{opt}}$.

This instability is related to the feedback process discussed below Eq.\eqref{in-out detector} due to the reflection of the SRM. Intuitively, when the reflectivity of SRM $r_s$ becomes high (or equivalently, $t_s$ decreases), the photon loss rate through the transmission for each round trip can be less than
the photon increasing rate through amplification by the gain medium, corresponding to the \textbf{``optical instability"}. The criterion of this instability is determined by the analytical behavior of the close-loop transfer matrix $\textbf{M}_c$. In our configuration, this close-loop transfer matrix is diagonal so that it can be simplified as a close-loop transfer function $G_c(\Omega)$ ($\textbf{M}_c=G_c(\Omega)\textbf{I}$):
\be\label{close loop gain}
G_c(\Omega)=\frac{1}{1-r_sG_o(\Omega)}
\ee
with $G_o(\Omega)=e^{2i\Omega\tau}M(\Omega)$.
The stability of the full system is determined by the poles of the denominator, which can be obtained by solving the equation $1-r_sG_o(\Omega)=0$.

\begin{figure}
  \includegraphics[width=0.4\textwidth]{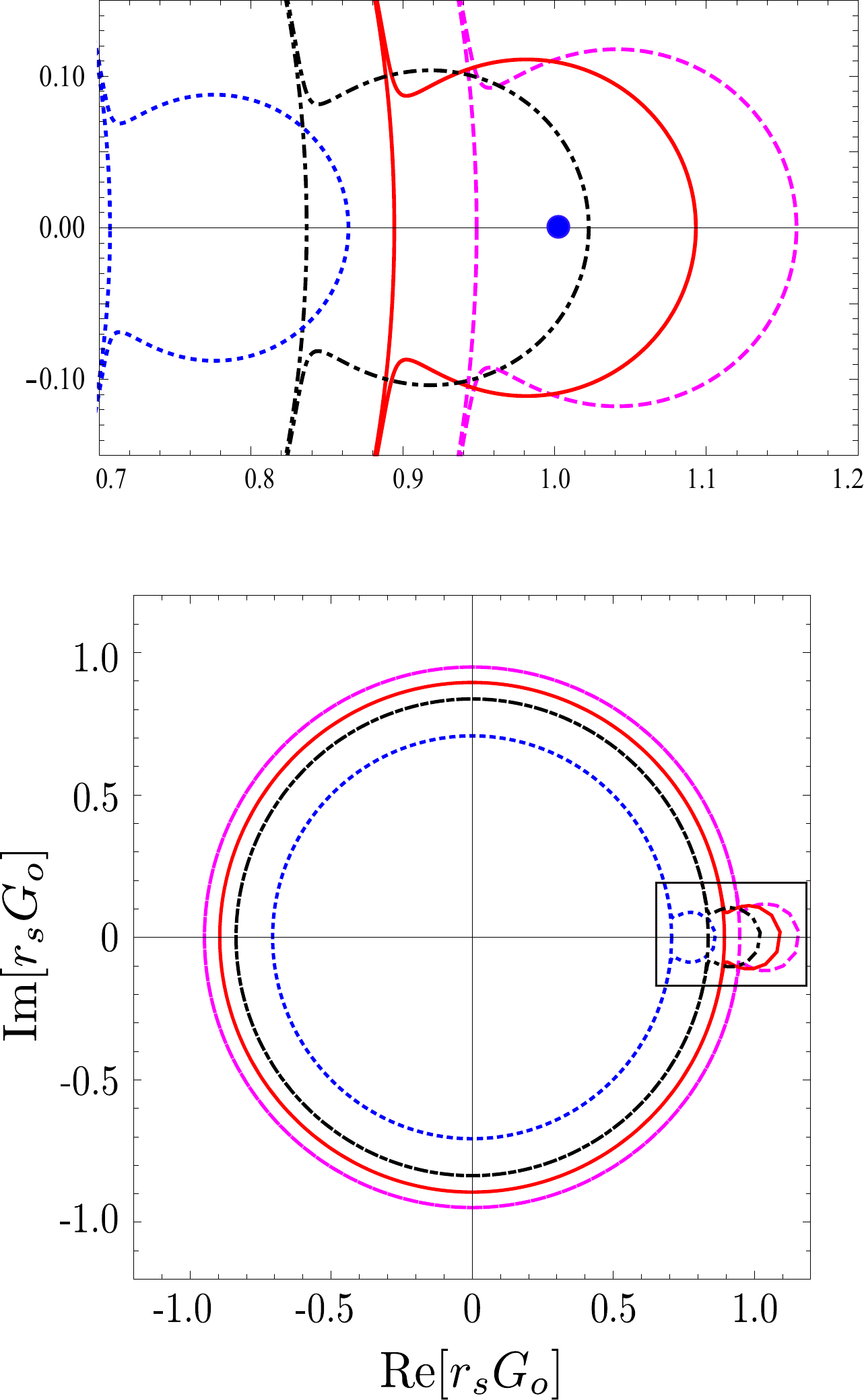}\\
  \caption{Nyquist contours of the $r_sG_o(\Omega)$ for the full system with fixed parameters of the gain medium $\eta=0.1,\xi=0.4$ while varying the SRM amplitude reflectivity $r_s$. The dashed (magenta), solid (red), dotdashed (black), dotted (blue) curves are the Nyquist contours when $r_s^2=0.9, 0.8, 0.7 \text{ (unstable cases), and } 0.5 \text{ (stable cases) }$, respectively. The upper-part zooms-in the full contour (lower-part) near (1,0) (the red spot). It is clear here that when the SRM reflectivity increases, the instability develops and the stable region in \figref{stability} shrinks.}\label{Nyquist}
\end{figure}

However, the time-elapsed factor $e^{2i\Omega\tau}$ in the gain function makes it difficult to find the root of the above mentioned equation. The \textbf{Nyquist criteria} provides us another way to understand the stability through the analytical behavior of $G_o(\Omega)$\,\cite{Nyquist}(See \emph{Appendix} C) instead of $G_c(\Omega)$. Specifically in our system, the Nyquist criteria can be stated in a way that the Nyquist contour of $r_sG_o(\Omega)$ should \textbf{not} encircle the point $(1,0)$ in the $(\text{Re}[r_sG_o],\text{Im}[r_sG_o])$ plane at all. This criteria is equivalent to the lasing condition that the round-trip gain is smaller than one when the phase is the integer number of $2\pi$. For illustration purpose, several examples of the Nyquist contour of $r_sG_o(\Omega)$ are shown in \figref{Nyquist} given typical parameters of $\gamma_{12}$ and $\Delta_0$.
This plot demonstrates that increasing the SRM reflectivity can lead to system instability. We further search the parameter region $0<\eta,\xi<1$ and give the plot on Fig.\ref{stability}, from which we can see that the stability criteria imposes a very strong constrain on the possible parameter region. Only the interferometer system with parameters in the stable region is useful.

\subsection{Integrated shot noise limited sensitivity improvement factor}
For quantitatively describing the improvement of the integrated shot noise limited sensitivity, we define a quantity: the integrated shot noise limited sensitivity improvement factor (iSNS improvement factor) $\rho_r$ in the following way:
\be\label{SNRg}
\rho_r=\int^{\omega_{\text{FSR}}}_0\frac{1}{S^a_{hh}(\Omega)}d\Omega/\int^{\omega_{\text{FSR}}}_0\frac{1}{S_{hh}(\Omega)}d\Omega,
\ee
where $S^a_{hh}$ is the shot noise limited gravitational wave strain sensitivity of the laser interferometer with double-pumped gain medium given by Eq.\eqref{sensitivity}, while the denominator is that of the conventional interferometers given in Eq.\eqref{limit}.
In case of $\rho_r>1$, the system with double gain media will improves the signal to noise ratio by breaking the trade-off between the detection bandwidth and peak sensitivity. However, we need to be cautious about the stability of the system at the same time.
\begin{figure}
  \includegraphics[width=0.5\textwidth]{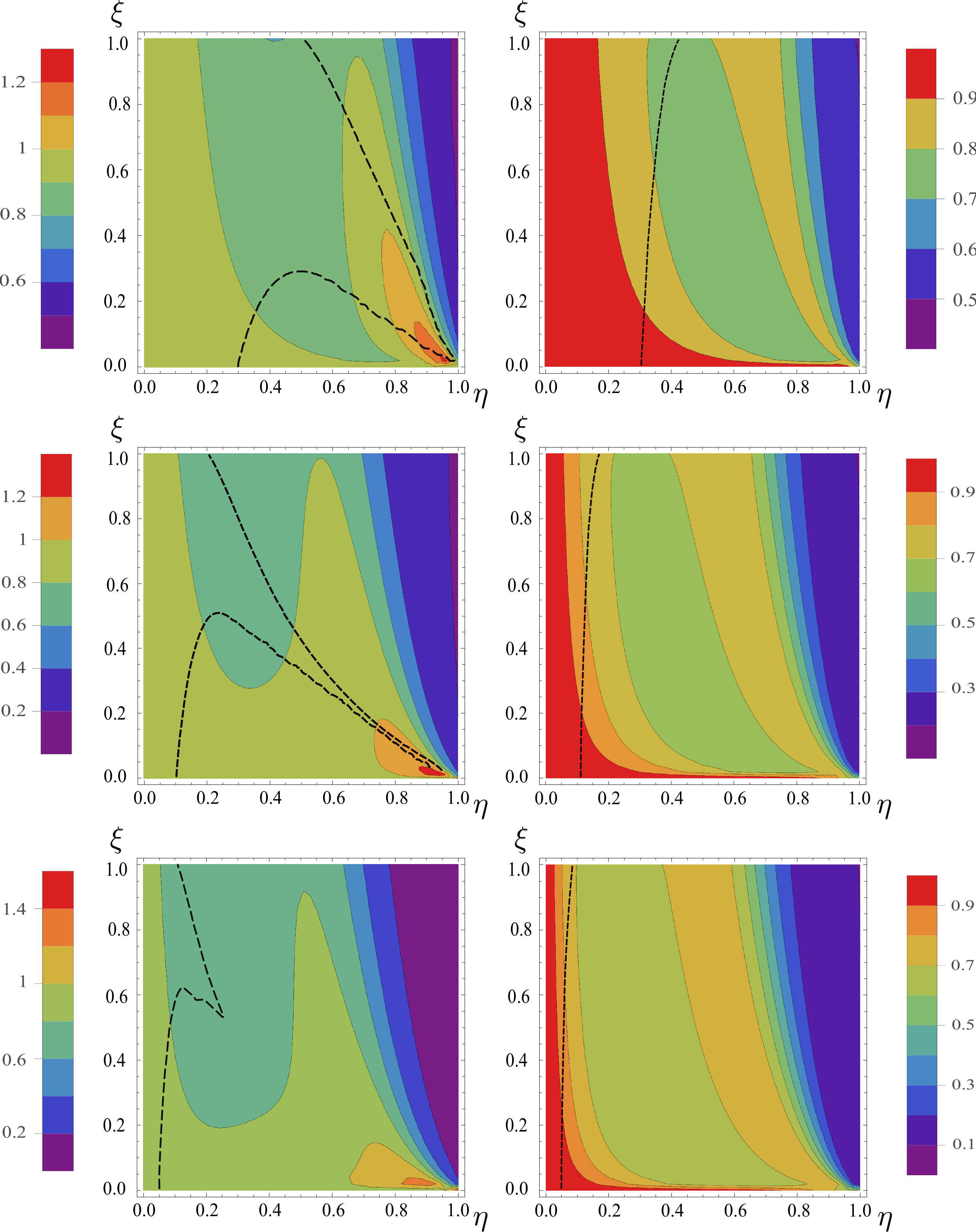}\\
  \caption{Integrated shot noise limited sensitivity improvement factor (defined in Eq.\eqref{SNRg}) of the full interferometer scheme with double gain medium, without the effect of additional noise. The specification for the parameters is identical to the one for producing \figref{stability} and \figref{gain_noise}. The left panel and right panel correspond to the larger and smaller roots of Eq.\eqref{PCcondition}, respectively. The dashed line is the boundary of the stable region shown in \figref{stability}. In this figure, when the detuning takes the larger solution, there are some regions where $\rho>1$ and the system is stable at the same time. However, as we can see from \figref{gain_noise}, these regions will disappear when we take into account of the effect of the additional noise.}\label{sensitivity_nonoise}
\end{figure}

We can calculate the strain sensitivity and hence the iSNS improvement factor. By fixing the SRM power reflectivity to be $r_s^2=0.8$, we calculate the iSNS improvement factor by searching the parameter region for $(\eta,\xi)$ within the range $[0,1]$ constrained by the phase cancelation condition.
For illustration purpose, we first calculate the $\rho_r$ by ignoring the effect of additional noise introduced by the atom system and give the plot in \figref{sensitivity_nonoise}. This figure clearly shows that there could be some parameter regions where the system is stable and $\rho>1$.

However, taking into account of the additional noise,  the results dramatically changed as we can see from \figref{gain_noise}. It turns out that there is no region where $\rho>1$ and the $\rho-$ contours are significantly distorted due to the additional noise. According to our numerical test,
this conclusion does not change with the variation of the SRM reflectivity.


\section{Conclusion}
In this paper, we applied the input-output formalism developed from the Hamiltonian of light-atom interaction to study the quantum noise of white light cavity using double gain medium. We find that not only does the additional noise associated with the parametric amplification process affects the system, but the requirement for the system stability also introduces an additional issue to take into account for its implementation. We conclude that the net sensitivity can not be enhanced by using the anomalous dispersive behavior of the stable double gain medium when the system is stable. For further study, we will consider the situation that the system is unstable but being controlled by a external feedback loop in an accompanied paper.

{\it Acknowledgements---}
We thank Atsushi Nishizawa, Bassam Helou, Belinda Pang, and other members of LIGO-MQM discussion group for fruitful discussions. We thank Vsevolod Ivanov for reading our manuscript. We also thank S. Shahriar for giving useful comments on the manuscript. Y.M is supported by the Australian Departement of Education, Science and Training. C.Z is supported by the Australian Research Council. H. M is supported by the Marie-Curie Fellowship. Y. C is supported
by NSF Grants PHY-1068881 and CAREER Grant PHY-0956189. Y. M would like 
to thank Li Ju and David Blair for their keen support of his visit to
Caltech where this work has been done. 

\appendix
\section{Slowly-varying amplitude Hamiltonian of the electromagnetic field}
In the main text, the Hamiltonian of the electromagnetic field is given in Eq.\eqref{free field Hamiltonian}. Unlike the usual free field Hamiltonian written in the $k-$space, this Hamiltonian is written in the $x-$space and the $\hat a_{x}$ is the slowly varying amplitude of the optical field. In this appendix, we give an derivation of this form of Hamiltonian.

In the $k-$space, the free-field Hamiltonian for an unidirectional propagating field can be written as:
\be
\hat H_{f}
=\hbar c\int^{\Delta k}_{-\Delta k}dk' (k_p-k')\hat a_{-k_p+k'}^\dag\hat a_{-k_p+k'},
\ee
Since we are interested in the $\Delta k\ll k_0$, we can approximate the above formula to be:
\be\label{EM_Hamiltonian_k}
\hat H_{f}\approx \hbar c\int^\infty_{-\infty}dk' (k_0-k')\hat a_{-k_0+k'}^\dag\hat a_{-k_0+k'}.
\ee
This approximation is called \emph{narrow band approximation}.

Then we can define the optical field operator in the $x-$space by Fourier transformation:
\be
\hat{\tilde{a}}_x\equiv\int^\infty_{-\infty}\hat a_{-k_0+k'}e^{-ik'x}dk'
\ee
Substituting the above definition into the Eq.\eqref{EM_Hamiltonian_k}, we obtain:
\be
\hat H_{f}=\hbar ck_0\int^\infty_{-\infty}dx\hat {\tilde{a}}_x^\dag\hat{\tilde{a}}_x+\frac{i\hbar c}{2}\int^\infty_{-\infty}dx\left[\frac{\partial \hat{\tilde{a}}_x^\dag}{\partial x}\hat{\tilde{a}}_x-\hat{\tilde{a}}_x^\dag\frac{\partial \hat{\tilde{a}}_x}{\partial x}\right]
\ee
Further more, if we work in the rotating frame of $ck_0$:
$\hat a_x=\hat{\tilde{a}}_x e^{-ick_0t}$ then the above Hamiltonian will be:
\be
\hat H_f=\frac{i\hbar c}{2}\int^\infty_{-\infty}dx\left[\frac{\partial \hat{\tilde{a}}_x^\dag}{\partial x}\hat{\tilde{a}}_x-\hat{\tilde{a}}_x^\dag\frac{\partial \hat{\tilde{a}}_x}{\partial x}\right]
\ee
Notice that the $\hat a_{x}$ satisfies the commutation relation: $[\hat a_x,\hat a^\dag_{x'}]=\delta(x-x')$.
The $\hat a_x$ here is the spatially and temporary slowly varying amplitude of the electromagnetic field. This fact can be seen from the definition of the electric field under the narrow-band approximation:
\be
\begin{split}
&\hat{\tilde{E}}^{(+)}(x,t)=\hat E^{(+)}(x,t)e^{i\omega_0t+ik_0x}\\&\approx\sqrt{2\pi \hbar ck_0}
\int^\Delta k_{-\Delta k}dk'\hat a_{-k_0-k'}e^{-ik'x-ick't}=\sqrt{2\pi \hbar ck_0}
\hat a_x
\end{split}
\ee

\section{Validity condition and Phase cancelation condition}
In the main text, we mentioned that $\chi(\Omega)$ given by Eq.\eqref{suseptibility} and the input-output relations Eq.\eqref{in-out} and Eq.\eqref{Mcoefficient}-\eqref{Ncollectivecoefficient} are obtained by the neglecting the terms with frequency $\sim 2\Delta_0$. Here, we are going to give a derivation of this term to show how it affects the field input-output relation of the double gain medium. The order of magnitude of this effect determines the validity region of Eq.\eqref{trans}. Notice that for simplicity, we only discuss the case of symmetrical pumping with $E_a=E_b=E_c$.

We start from the single-atom dynamics. As we can see in the main text, the equation of motion for $\hat\sigma_{12}$ after the adiabatic elimination of $\hat\sigma_{23}$ can be written as:
\be\label{sigma12a}
\begin{split}
\dot{\hat\sigma}_{12}+(\gamma_{12}-\gamma_{\text{opt}})\hat\sigma_{12}=&\frac{i}{2}\mu_{23}\hat a^\dagger_{\rm in}\bar{\sigma}_{13}+\gamma^s_{2\Delta_0}\hat\sigma_{12}\\
&-\sqrt{2\gamma_{12}}\hat n_{12\rm in}.
\end{split}
\ee
Here for simplicity, the pumping strength is assumed to be symmetric:
\be
\bar\sigma_{13}(t)\approx\frac{\mu_{13}E_c}
{\omega_{31}-\omega_{0}}
\cos(\Delta_0t).
\ee

Assume that $\hat\sigma_{12}\rightarrow\hat\sigma^{(0)}_{12}+\hat\sigma^{\text{ad}}_{12}$ where $\hat\sigma^{(0)}_{12}$ is the part which satisfies the Eq.\eqref{sigma12a} without the $\gamma_{2\Delta_0}$ term and $\hat\sigma^{\text{ad}}_{12}$ is due to the effect of $\gamma_{2\Delta_0}$ term. Then using iteration method, $\hat \sigma^{\text{ad}}_{12}$ satisfies (temporarily neglecting the $\hat n_{12}$ term):
\be\label{sigma12add}
\dot{\sigma}^{\text{ad}}_{12}+(\gamma_{12}-\gamma_{\text{opt}})\sigma^{\text{ad}}_{12}
=\gamma^s_{2\Delta_0}\hat\sigma^{(0)}_{12}.
\ee
The $\hat\sigma^{(0)}_{12}$ in the above equation can be solved as:
\be
\hat\sigma^{(0)}_{12}(t)\approx2i\sqrt{\gamma_{\text{opt}}}
\int^t_{-\infty}e^{(\gamma_{12}-\gamma_{\text{opt}})(t'-t)}\cos(\Delta_0t')
\hat a^\dagger_{\text{in}}(t')dt',
\ee
which leads to:
\be
\begin{split}
\gamma_{2\Delta_0}\hat\sigma^{(0)}_{12}(t)=&
\frac{i}{2}\gamma_{\rm opt}\sqrt{\gamma_{\text{opt}}}
\int d\Omega \hat a^\dagger_{\text{in}}(\Omega)\left[\frac{e^{-i(\Delta_0+\Omega)t}}{\gamma_{12}-\gamma_{\text{opt}}+i(\Delta_0-\Omega)}\right.
\\&\left.+\frac{e^{i(\Delta_0-\Omega)t}}{\gamma_{12}-\gamma_{\text{opt}}-i(\Delta_0+\Omega)}\right]
+O(3\Delta_0).
\end{split}
\ee
The terms oscillating around $3\Delta_0$ will be neglected in the further discussion. Substituting the above formula into Eq.\eqref{sigma12add}, we obtain $\hat \sigma^{\text{add}}_{12}$ as:
\be
\begin{split}
\hat \sigma^{\text{add}}_{12}(\Omega)
=&\frac{i\sqrt{\gamma_{\rm opt}}\gamma_{\rm opt}}{2(\gamma_{12}-\gamma_{\rm opt}-i\Omega)}
\left[\frac{\hat a^\dag_{\rm in}(\Omega+\Delta_0)}{\gamma_{12}-\gamma_{\rm opt}-i(\Omega+2\Delta_0)}\right.\\&\left.+\frac{\hat a^\dag_{\rm in}(\Omega-\Delta_0)}{\gamma_{12}-\gamma_{\rm opt}-i(\Omega-2\Delta_0)}\right].
\end{split}
\ee
Based on this equation, using the relation Eqs.[\ref{a-bsigma23}] and [\ref{sigma23}], we have the input-output relation as:
\be
\begin{split}
M(\Omega)=&1+f_-(\Omega)\frac{\gamma_{\text{opt}}}{-i\Omega-i\Delta_0+\gamma_{12}-\gamma_{\text{opt}}}\\
&+f_+(\Omega)\frac{\gamma_{\text{opt}}}{-i\Omega+i\Delta_0+\gamma_{12}-\gamma_{\text{opt}}}.
\end{split}
\ee
in which
\be
f_{\pm}(\Omega)=1+\frac{1}{2}\frac{\gamma_{\text{opt}}}
{\gamma_{12}-\gamma_{\text{opt}}+i(\Omega\pm\Delta_0)},
\ee
where $\gamma^s_{\text{opt}}$ is given by Eq.\eqref{gammaopt} of the main text.
Then the validity condition for our input-output relation is clear:
\be
\left|f_{\pm}-1\right|^2\ll1.
\ee
In the balance pumping case and under the D-C assumption $\Omega\sim0$, we can simplify the above validity condition to be:
\be\label{validity}
\Delta_0^2+(\gamma_{12}-\gamma_{\text{opt}})^2\gg(\gamma_{\text{opt}})^2/4.
\ee

In the case of many-atoms case, the $\gamma^s_{\rm opt}$ in the above validity condition will be replaced by $\Gamma_{\rm opt}$.

If the validity condition for the input-output relation was satisfied, the input-output relation Eq.\eqref{Mcoefficient}-\eqref{Ncollectivecoefficient}  can be used in analyzing the system. Moreover,
as we have mentioned in Section I, this condition is equivalent to the weak coupling approximation which allows us to approximate the phase generated by the gain medium to be ${\rm Re}[\chi(\Omega)]/2$. Besides, for achieving cancelation of the propagating phase inside the arm cavity in the weak coupling limit, another condition must be satisfied:
\be\label{phasecancelation}
\frac{\Gamma_{\text{opt}}[(\gamma_{12}-\Gamma_{\text{opt}})^2
-\Delta^2_0]}{[(\gamma_{12}-\Gamma_{\text{opt}})^2
-\Delta^2_0]^2}=-\frac{L_{\rm arm}}{c}.
\ee
This condition will reduce to $\Gamma_{\text{opt}}=\Delta_0^2L_{\rm arm}/c$ when $|\gamma_{12}-\Gamma_{\text{opt}}|\ll\Delta_0^2$. In our calculation, we have used the exact formula Eq.\eqref{phasecancelation}.

If we fixed the value of $\gamma_{12}$ and $\Gamma_{\text{opt}}$, then the phase cancelation condition becomes a second order algebraic equation for $\Delta_0^2$. Suppose this equation has two roots $x_1,x_2$, then we have:
\be
\begin{split}
&x_1x_2=(\gamma_{12}-\Gamma_{\text{opt}})^2[(\gamma_{12}-\Gamma_{\text{opt}})^2+\Gamma_{\text{opt}}c/(2L_{\rm arm})],\\
&x_1+x_2=\Gamma_{\text{opt}}c/(2L_{\rm arm})-2(\gamma_{12}-\Gamma_{\text{opt}})^2.
\end{split}
\ee
Notice that in the above equations, $x_1x_2$ is always positive, thereby $x_1+x_2$ can only be positive:
\be\label{x1+x2}
(\gamma_{12}-\Gamma_{\text{opt}})^2<\Gamma_{\text{opt}}c/(4L_{\rm arm}).
\ee
On the other hand, Eq.\eqref{phasecancelation} must have real roots, which gives:
\be\label{criteria}
(\gamma_{12}-\Gamma_{\text{opt}})^2<\Gamma_{\text{opt}}c/(8L_{\rm arm}),
\ee
which is a more stringent condition than Eq.\eqref{x1+x2}.

In summary, considering the validity of our input-output relation Eq.\eqref{Mcoefficient}-\eqref{Ncollectivecoefficient} and the requirement of the phase-cancelation condition, our parameters must satisfy Eq.\eqref{validity} and Eq.\eqref{criteria}. It is important to notice that there are always two $\Delta_0^2$ corresponding to a fixed set of $(\gamma_{12},\Gamma_{\text{opt}})$. In plotting the \figref{gain_noise} and \figref{sensitivity_nonoise}, we should take into account both roots.

\section{Nyquist criteria}
In Appendix C, we will give a brief introduction to Nyquist criteria\,\cite{Nyquist} which we used in understanding the stability condition of the full system in Section IV.

The behavior of control systems is usually described by gain functions. For a control system with feedback process, the open-loop gain function $G_o(\Omega)$ is used to describe the information transfer ignoring the feedback process, while
the closed loop gain function $G_c(\Omega)$ includes the effect of the feedback process. The relationship between $G_o(\Omega)$ and $G_c(\Omega)$ can be written as:
\be
G_c(\Omega)=\frac{G_o(\Omega)}{1+H(\Omega)G_o(\Omega)}.
\ee
The $H(\Omega)$ is the gain function for the feedback process itself, it is clear from \figref{flow chart} that in our system, it is just the reflection of the SRM: $-r_s$.

The stability of the system depends critically on the poles of the close-loop transfer function, that is, it depends on the poles of $G_o(\Omega)$ and also the zeros of $1-r_sG_o(\Omega)$. However, computing the poles and zeros of these gain functions is generally a difficult task when these functions are non-rational. The Nyquist stability criterion is a graphical technique for determining the stability of a control system, which is based on the following \emph{Lemma}: Cauchy argument principle.

The Cauchy argument principle starts from the Nyquist mapping, which maps the complex argument $\Omega$-plane to the complex  $F(\Omega)$-plane. If we have a clockwise contour in $\Omega$-plane encircling a \emph{zero} of $F(\Omega)$, correspondingly, the contour also encircles the \emph{origin} \emph{clock-wisely} in the $F(\Omega)-$plane. However, if we have a clockwise contour in $\Omega$-plane encircling a \emph{pole}, then the corresponding contour will encircles the \emph{infinity}\emph{ clock-wisely} in the $F(\Omega)-$plane thereby encircling the \emph{origin} in an \emph{anti-clockwise} way. In general, if we have a contour in the $F(\Omega)$-plane encircling the origin $N$ times clock-wisely, that means in the $\Omega-$plane, the number of zeros ($Z$) and the number of poles ($P$) satisfy:
\be
Z=N+P.
\ee
This equality is the Cauchy argument principle.

The transformation for quantity $A(t)$ between the frequency domain and the time domain is defined as: $A(t)=\int^{\infty}_{-\infty}A(\Omega)e^{-i\Omega t}$. Therefore if $A(\Omega)$ has poles in the upper-half plane, we will have instabilities for casual system ($t>0$). Now we choose the contour encircling the upper-half $\Omega-$plane as ``Nyqusit contour". If the system is stable, then the $Z$ of $1-r_sG_o(\Omega)$ (the denominator of close-loop gain function) inside the Nyquist contour should be zero. As a result, the Cauchy argument principle becomes $N=-P$, which is the Nyquist criteria.

In our system with $G_o(\Omega)=\chi(\Omega)e^{2i\Omega\tau}$, we have the the poles of $r_sG_o(\Omega)$ which are $\Omega_{1,2}=\pm\Delta_0-i(\gamma_{12}-\Gamma_{\text{opt}})$. Both of them fall outside of Nyquist contour because $\gamma_{12}-\Gamma_{\text{opt}}>0$ for the requirement of the stability of the atomic gain medium itself. Then we can conclude that $P=0$ inside the Nyquist contour. In this case, the Nyquist criteria requires $N=0$ to keep the stability for the full system, that is, in the Nyquist diagram, the contour of $1-r_sG_o(\Omega)$ should \emph{not } encircle the origin \emph{at all}. In other words, the contour of $r_sG_o(\Omega)$ should not encircle the point $(1,0)$ in the $(\text{Re}[r_sG_o(\Omega)],\text{Im}[r_sG_o(\Omega)])$ plane.


\end{document}